# Data Driven Authentication: On the Effectiveness of User Behaviour Modelling with Mobile Device Sensors

Hilmi Güneş Kayacık*, Mike Just*, Lynne Baillie*, David Aspinall† and Nicholas Micallef*
*Glasgow Caledonian University, Glasgow, UK, {gunes.kayacik, mike.just, lynne.baillie, nicholas.micallef}@gcu.ac.uk
†University of Edinburgh, Edinburgh, UK, david.aspinall@ed.ac.uk

*Abstract*—We propose a lightweight, and temporally and spatially aware user behaviour modelling technique for sensor-based authentication. Operating in the background, our data driven technique compares current behaviour with a user profile. If the behaviour deviates sufficiently from the established norm, actions such as explicit authentication can be triggered. To support a quick and lightweight deployment, our solution automatically switches from training mode to deployment mode when the user's behaviour is sufficiently learned. Furthermore, it allows the device to automatically determine a suitable detection threshold. We use our model to investigate practical aspects of sensor-based authentication by applying it to three publicly available data sets, computing expected times for training duration and behaviour drift. We also test our model with scenarios involving an attacker with varying knowledge and capabilities.

## I. INTRODUCTION

Mobile devices such as smart phones and tablets are rapidly becoming our digital identity. They are used for payments and authentication, and they store valuable information. Today's mobile device hardware is quite capable with multi-core gigahertz processors, and gigabytes of memory and solid-state storage. Their relatively low cost, ease of use and 'always on' connectivity provides a suitable platform for many day-to-day tasks involving money and sensitive data, which in turn makes mobile devices an attractive attack target (e.g., see attacks against well-known Apple iOS and Google Android platforms [1]).

Authentication based on user behaviour and biometrics has attracted interest as mobile device popularity grows. For traditional authentication methods, research shows that PIN and password-based security is cumbersome to use [2], [3], [4] and is frequently disabled by users. A recent study [5] shows that 64% of users do not use authentication on their phones. By contrast, *implicit authentication* relies not on what the user knows but is based upon user behaviour, and is accomplished by building so-called *user profiles* from various sensor data [3], [6], [7], [8], [9], [10]. If the user behaviour is consistent with their profile, the device will have high *comfort*, hence no explicit authentication action is required. However, if the user deviates sufficiently from the established normal behaviour, alternative measures can be triggered, such as requiring a PIN or password. Reducing the occurence of explicit authentication should provide a more usable experience, and might encourage more users to protect access to their devices.

Generally, previous work has ignored the practical considerations of deploying implicit authentication in large scale, in-the-wild deployment scenarios. In this work, we provide a data driven and empirical study of implicit authentication under a realistic scenario. In this case, we assume that the user purchases a mobile device with behaviour learning capabilities and starts using the device in the same way that they would any other just-purchased mobile device. The device starts in 'training mode', learning the user's routine until it determines *automatically*, from data-driven heuristics that it is time to switch to deployment. When the model is built and training ends, the device is considered to be in 'deployment mode' in which it presents authentication challenges when a user deviates from the established normal behaviour. Our research sought to answer the question of whether we could build and deploy such a model in a practical and secure manner on today's mobile devices, in support of which this paper contributes the following:

1) We propose an *incremental training paradigm* that is transparent to the user. In training mode, the device updates the profile daily. When the device learns a user's routine sufficiently, it automatically switches from training to deployment mode.
2) After training, the device automatically determines a suitable *detection threshold*, below which explicit authentication is invoked. As the user interacts with the device day after day, the threshold is dynamically updated from the observed data.
3) In recognition of changes in work patterns, travel to new locations and moving to a new place, we present case studies where *behavioural drift* occurs and examine different retraining techniques.
4) We also present an *attack analysis*, based upon an adversary with varying knowledge and capabilities.

Our incremental training differs from the previous work [3], [11], [12] which uses a fixed subset of data (i.e. a percentage of collected data, typically over a few weeks, is used for training). We do not disagree that a few weeks may suffice for training, in fact, we empirically confirm it. We however argue that training duration must be set automatically on a per user basis since, as our evaluation shows, there is no one-size-fits-all.

The remainder of the paper is organised as follows. Related mobile device sensing work is discussed in Section II. The datasets used in our experiments are detailed in Section III. Our methodology for building user profiles is introduced in Section IV, including our incremental training approach and automatic threshold generation. Section V provides the results, including our analysis of behavioural drift and our attack scenarios. Conclusions are drawn in Section VI.

## II. RELATED WORK

Building profiles from mobile device sensors has a broad range of possible applications, for example, to facilitate social studies of human behaviour [6], [13], [14], publicly available data collections [15], [16] and context aware devices [17], [3], [11], [8], [9], which is the focus of this work.

Similar to our work, Gupta et al. [17] proposed a model for the familiarity and safety of a user's device based upon its location, and used this to automatically construct access control policies. Their model distinguishes the behaviour of different users, and incorporates user feedback for refinement, though they do not consider the duration of training or the transition from a training to deployment or retraining. Shi et al. [3] focused on implicit authentication by learning user behaviour and assigning a score – positive for familiar events and negative for novel – based on recent user activity. The training was performed on a fixed subset (60%) of the data. Lin et al. [7] proposed a non-intrusive authentication method based on orientation sensor data using k-nearest neighbour classification. They argued that while input from a single sensor may yield poor accuracy, combining multiple sensor inputs would improve the accuracy. To this end, Senguard [18] aimed to implicitly and continuously authenticate users using input from many sensors yielding a stronger classifier built from per-sensor classifiers.

Furthermore, context aware authentication research [4], [8], [9], [10] focused on sensing the context in which the device is used (such as home or work) and providing access based on device comfort computed from various sensor data. Though a location-only may be more susceptible to insider attacks (e.g., friends and family), something that we address in our attack model. Eagle et al. [6] employed Eigenbehaviour analysis to identify the patterns in a user's daily routine. Using MIT Reality Mining Data [15], a small set of characteristic vectors are computed, summarising user behaviour. Various research [3], [11], [12] proposed an implicit authentication method using behavioural and environmental biometrics collected from device sensors. While this context-aware, sensor-based authentication research is similar to our work, the do not address the incremental training and determining detection threshold automatically from data.

While previous work investigated the use of various modelling methods, very little attention was paid to the practical considerations of using such methods in large-scale deployments with minimum intervention. It is reasonable to expect a (motivated) user to expend some effort in 'teaching' the device by providing feedback but they will quickly grow tired if frequent and labour intensive feedback is required. Additionally, some users may prefer an on-device modelling technique that allows them to build and deploy models without their data ever leaving the device. Thus, in this work, we propose a lightweight, non-parametric modelling approach that can run on today's modern devices and determine when to stop training and the threshold for detection, both automatically from the data. It is conceivable to have multiple techniques, some in the cloud for more intensive but also accurate modelling. Thus we believe our technique complements the existing work by providing a scalable alternative that can run on a device.

## III. DATASETS

To facilitate our analysis we use the publicly available Rice and MIT datasets, as well as our own GCU dataset. Table I summarises the datasets used in our analysis.

### A. GCU

The GCU dataset currently consists of a collection from 7 staff and students of Glasgow Caledonian University. The present data was collected in 2013 from Android devices and contains sensor data from wifi networks, cell towers, application use, light and noise levels and device system stats. The duration of the data varies from 2 weeks to 14 weeks for different users. Compared to other publicly available datasets used in this paper, it also contains a detailed diary for each user which allows for a more informed investigation of anomalies. The dataset is publicly available in text file format[1].

### B. Rice Livelab

The Rice Livelab dataset [16] was created from the behaviour of 35 users, all students at Rice University or Houston Community College. The data was collected from iPhone 3GS devices between 2010 and 2011 and contains sensor data such as application use, wifi networks, cell towers, GPS readings, battery usage and accelerometer output. The duration of the data varies from a few days to less than one year for different users. The dataset is publicly available in MySQL format.

### C. MIT Reality Mining

The MIT Reality Mining dataset [15] contains the behaviour data of 100 subjects from various departments of MIT. The data was collected from Nokia 6600 smartphones between 2004 and 2005 and contains sensor data such as call logs, bluetooth devices in proximity, cell towers, application usage. The duration of data collection varies from a few days to about one year for different users. It is publicly available as a MATLAB workspace file.

TABLE I. SUMMARY OF THE GCU, RICE AND MIT DATASETS. COLLECTION YEAR IS PROVIDED IN PARENTHESES.

|  | **GCU (2013)** | **Rice (2010)** | **MIT (2004)** |
|---|---|---|---|
| **Users** | 7 | 25 | 100 |
| **Duration** | 3 weeks | 12 months | 6 months |
| **Sensors** | app, wifi, cell, cpu load, light, noise, magnetic field, rotation | app, wifi, cell, device active, call history, battery, cpu load | app, bluetooth, cell, device active, call, charge |

## IV. METHODOLOGY

Our profiling technique builds temporal and spatial models from data in a lightweight and non-parametric way. When the profile stabilises, training is considered to be complete and the device switches to a deployment mode. A detection threshold is computed based on the user's security settings to activate explicit authentication when the comfort is below the threshold.

---

[1]Partial data can be accessed at http://www.ittgroup.org/.

## A. User profile

Our technique builds user profiles from probability density functions of sensor data. Time and location from cell towers are generally present and are used as so-called *anchors*, though our model would support other sources such as wifi or GPS. Each anchor focuses on a specific location or time and describes the general characteristics for that anchor through a set of probability density functions for each sensor's data. Using the temporal and spatial models described below, the device can assign a comfort score for each sensor event based on its frequency of occurrence for that anchor, i.e., location or time. Scores from temporal and spatial models are aggregated, similar to [17], to produce the final score. This information allows the device to detect unusual use whether it is based on location or time (as detailed in Section V-C).

The time stamps from the cell towers allow for events to be partitioned, for example into 24 slots for each hour of the day. The location identifier from the cell tower provides an abstract label for a user's whereabouts in contrast to a GPS location that provides an exact location albeit with higher cost to the battery life [18]. Figure 1 provides an overview of the user behaviour model. In the temporal model there are 24 anchors, one for each hour of the day. On the other hand, the spatial model may contain as many locations as the user visits[2].

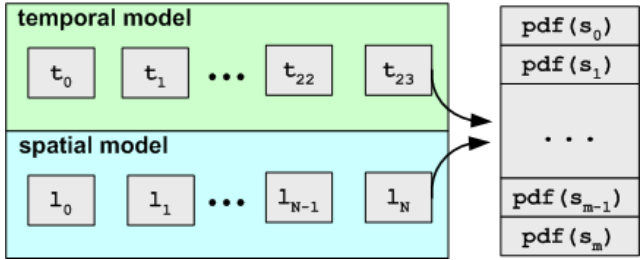

Fig. 1. Sample user behaviour profile that consists of temporal and spatial models. Each location and time-of-day is a set of probability density functions.

The number of sensors per anchor varies since for a location, or at a particular time, there may not be sufficient data for that sensor. For example, a location may not have any wifi networks associated with it or there may be no app use for a given time. This allows the model to opportunistically build the probability density functions for only the existing sensors. Thus, we define the temporal ($temp$) and spatial ($spat$) models as in Equations 1 and 2, where $L$ and $T$ denote location and time anchors. Each anchor has its own set of sensors $S = \{s_1, s_2, \ldots, s_m\}$ and $pdf$ denotes the probability density function.

$$model_{temp} = [pdf(s_1|T), pdf(s_2|T), \ldots, pdf(s_m|T)] \quad (1)$$

$$model_{spat} = [pdf(s_1|L), pdf(s_2|L), \ldots, pdf(s_m|L)] \quad (2)$$

Probability density functions for discrete sensors (e.g., wifi readings and app usage, where each data belongs to one category) are built using histograms [19] by maintaining the occurrence counts for each sensor event (e.g., using Facebook app at 3pm or checking email at home). Probability density functions for continuous data (e.g., light levels, cpu usage) are built using kernel density estimators (KDEs). KDE [20] is a non parametric way of computing probability distributions from continuous data.

---

[2]If a cell tower is not present, the location identifier is set to 'unknown.'

Building probability density functions has two distinct advantages. First, they can be built incrementally from the streaming data, thus training data does not need to be stored in memory, unlike k-nearest neighbour classification [7], support vector machine [9] approaches. Second, they are lightweight with typical computational complexity of $O(n)$ for discrete data of $n$ sample points and $O(n+m)$ for continuous data with $m$ evaluation points and $n$ sample points from the density. Figure 2 provides a short example of a user behaviour profile for a discrete and continuous set of sensor data. For our studies, a typical profile contains about 6 to 8 sensors.

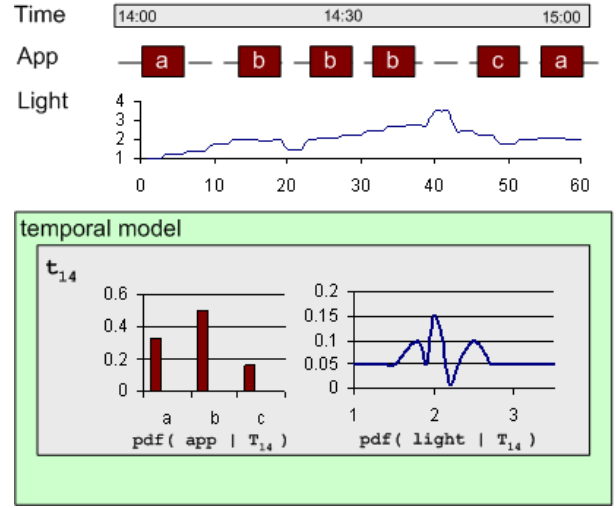

Fig. 2. Sample user behaviour profile that consists of temporal and spatial models for app usage and light readings. Each location and time-of-day is a set of probability density functions.

Each model focuses on establishing familiarity from different perspectives. For example, while the use of an application may be deemed frequent (hence, familiar) per location, if the use takes place at an unusual hour, it can be detected as a temporal anomaly. Similarly, if an application is used at a usual time but at an unusual location, it can be detected as a location anomaly. Maintaining multiple models, each with its own perspective allows us to build comfort levels based on multiple indicators. The device is more likely to become uncomfortable if the events are taking place at unusual times and locations. Similarly, as the device encounters familiar surroundings and behaviour, it will become more comfortable.

## B. Measuring comfort

A user profile consists of temporal and spatial models containing a set of probability density functions. These models are employed to establish comfort. The comfort – or lack thereof – is computed over a period of time and can be computed during training in order to measure profile stability (see Section IV-C) and during deployment in order to determine whether the device is comfortable or not (where the latter would trigger additional authentication actions).

Data from all available sensors are compared against the temporal and spatial models and each produces its own comfort score based on its frequency for the given time and location. If the location is unknown or the sensor data has never been encountered for that location, comfort is 0. Thus, each sensor input collected over the given time period provides a comfort score; note that some sensors may provide more than one input (e.g., noise and light readings are sampled more than once). The aggregate comfort score is computed as follows. The score from each sensor is aggregated into a sensor score first (Equation 3). It takes $n$ inputs from one sensor and compares it against the profile models based on the time and location which defines the anchor $A_{mod}$.

$$Score_{sen} = \frac{\sum_{i=1}^{n} pdf(sen_i | A_{mod})}{n} \quad (3)$$

Second, scores from many sensors for the given time and location are aggregated into temporal and spatial scores in Equation 4, where $sen = \{app, wifi, light, noise, cpu, call...\}$ as detailed in Table I for different datasets.

$$Score_{mod} = \frac{\sum_{sen} Score_{sen}}{|sen|} \quad (4)$$

Third, the overall aggregated comfort score, which determines the detection decisions, is computed by aggregating temporal and spatial scores in Equation 5 where $mod = \{temporal, spatial\}$. The aggregated score includes positive comfort from temporal and spatial models and a negative time score $score_{time}$, which is 0 if the previous and current readings are within 1 minute (sampling rate) interval. It increases to 1 if the readings are more than 60 minutes apart ensuring that comfort level gradually decreases if the user does not interact with the device.

$$Score_{agg} = \frac{\sum_{mod} Score_{mod}}{2} - score_{time} \quad (5)$$

This layered aggregation is advantageous: when the comfort is low, it is possible to determine if it is a temporal or spatial anomaly. Furthermore, it is also possible to determine which sensors provided low comfort scores. In our work, all sensors are given equal weights but this can be adjusted.

While the computed score provides a basis for comfort, a single event alone does not provide the sufficient level of granularity for establishing comfort. For example, even if the application was never used at a given location or hourly, events leading to the anomalous application use may indicate familiarity based on connected wifi and cell towers. This way, if the user is at a familiar location as established by wifi and cell tower models, an anomalous use of an application will produce 'discomfort' proportionally. Needless to say, if the available wifi networks are unfamiliar, cell towers indicate an unknown location and the application use is anomalous, the device will exhibit the most discomfort.

To this end, comfort is computed over a period of time (we empirically selected 1 minute intervals for our analysis in Section V). Comfort is computed continuously and incrementally so that scores from events that occurred in the current time frame are combined to produce the current comfort score.

### C. Profile stability: deciding when to deploy

In order to establish a comfort level which can serve as an authentication scheme, the device needs to 'learn' the behaviour of the user for a given period of time. Typically, previous studies used a fraction of the available data for training and used the rest for testing, e.g., [3], [9]. Conversely, we approach training from a data driven angle and aim to let the device decide when the training is 'sufficient.' To investigate how the learning rates change as the device learns the user behaviour, we measure the *profile stability*. A profile stabilises when the majority of the items such as wifi networks, applications, cell towers in the models remain unchanged between a number of days. We selected two consecutive days for our analysis. A key assumption for our implicit authentication solution – as well as those of other researchers – is that user behaviour does indeed show some 'stability' after several days. To compute stability, we establish two distance metrics, one based on Levenshtein distance [21] on discrete probability distribution models and an Euclidean distance metric that compares the percentile characteristics of continuous probability distribution models. For the latter, we rely on the property that the distance between two curves increases as their difference between percentiles increase.

The profiles are built incrementally every day and we compare the current day's sensors with the previous day's profile and assign a distance metric that ranges between 0 and 1. One would expect the distance to decrease as the models converge and fewer changes are observed between two consecutive day's profiles. Figures 3, 4 and 5 show the mean distance between the current and the previous profile for all users for spatial and temporal models. Plots show the typical distance per dataset not for a single user. Distance is computed for each sensor and the global distance compares the models as a whole.

Distance measures for GCU data in Figures 3(a) and 3(b) indicate that it takes at least a week for the profiles to stabilise, i.e., distance between subsequent days to settle. Given that GCU data is collected over a relatively short duration (two weeks), we also analysed MIT and Rice data. In both datasets, we observe that it takes a few weeks for the profiles to converge. This confirms the rule-of-thumb that previous work employed by utilising a fraction of the dataset that is typically a few weeks in duration. However, we argue that for different users, longer or shorter training durations may be suitable hence the device should be able to determine when to switch from training to deployment. Table II shows that profile for different users converge at different rates. The convergence is considered (from the empirical analysis detailed in Figures 3(a) and 3(b)) to take place when distance is below 0.1.

TABLE II.  NUMBER OF DAYS IT TAKES FOR PROFILES TO CONVERGE FOR GCU USERS.

|        | Convergence (Global) | Convergence (Temporal) | Convergence (Spatial) |
|--------|----------------------|------------------------|-----------------------|
| User 1 | 9 days               | 9 days                 | 9 days                |
| User 2 | 10 days              | 8 days                 | 10 days               |
| User 3 | 3 days               | 9 days                 | 1 days                |
| User 4 | 9 days               | 7 days                 | 9 days                |
| User 5 | 9 days               | 8 days                 | 14 days               |
| User 6 | 9 days               | 5 days                 | 11 days               |
| User 7 | 6 days               | 6 days                 | 8 days                |

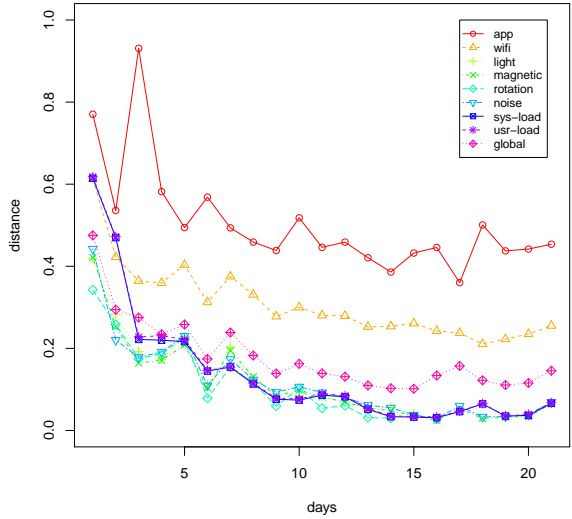
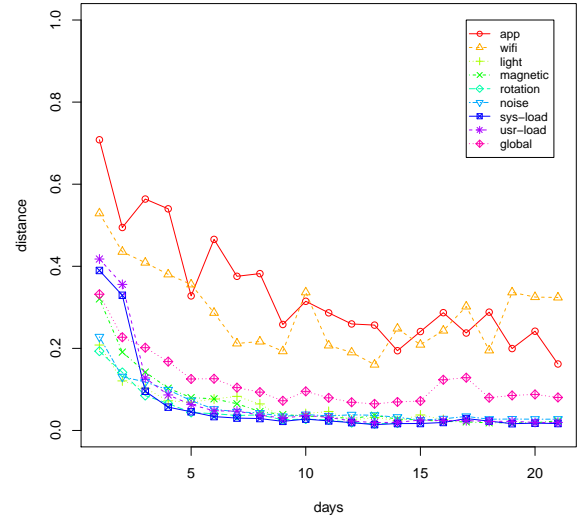

(a) Spatial models  (b) Temporal models

Fig. 3. Comparison of spatial and temporal models between current day and preceding day on GCU dataset. Low distance indicates increased similarity.

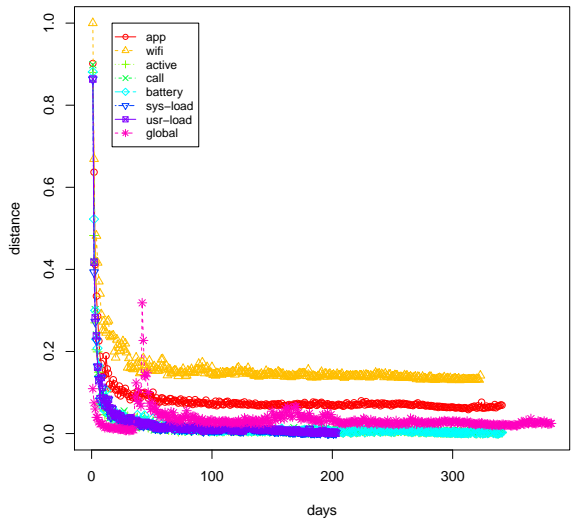
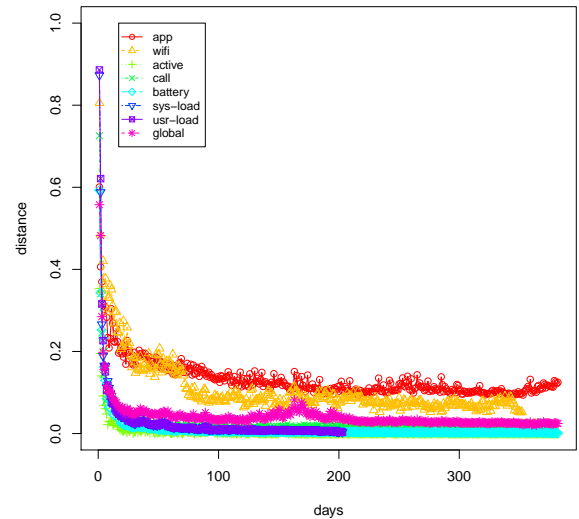

(a) Spatial models  (b) Temporal models

Fig. 4. Comparison of spatial and temporal models between current day and preceding day on Rice dataset. Low distance indicates increased similarity.

## D. Determining a detection threshold

Computing comfort scores as detailed in Section IV-B allows the device to produce a score at given time intervals, every 1 minute in this case. As profiles are built incrementally each day, the previous day's model is utilised to compare with the current day for convergence tests as shown in Section IV-C. The previous day's profiles can also be used to compute comfort levels for the current day's data to explore how device comfort changes over time. For example, one would expect the comfort to be low during the initial days since the device is unfamiliar with the user. As the profile converges and the training is sufficiently complete, the comfort should increase and the detection of anomalies now becomes possible. Each device will establish a comfort level unique to its user. While a daily average comfort level of 0.5 may be high for some users, for others (leading more predictable lives) it may be low.

A *detection threshold* can be defined as a value below which additional authentication must be performed, e.g., explicit authentication. We compute thresholds *per user* in a data driven fashion. We utilise the previous day's profile and compute comfort levels for the current day. The use of the previous day's profile allows us to investigate how the threshold might change during training. Needless to say, when a device is in deployment, the profile to be used will be

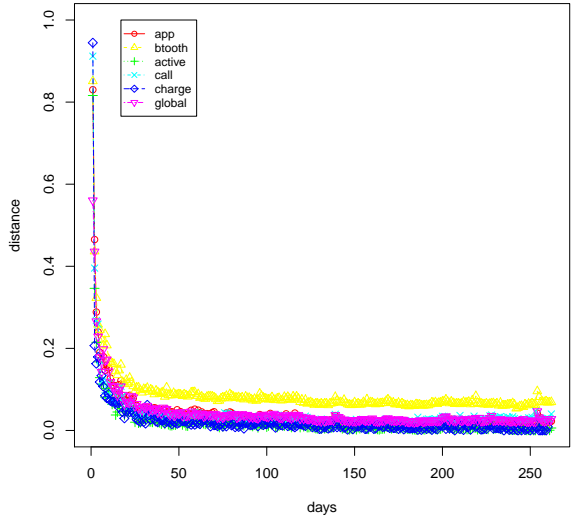
(a) Spatial models

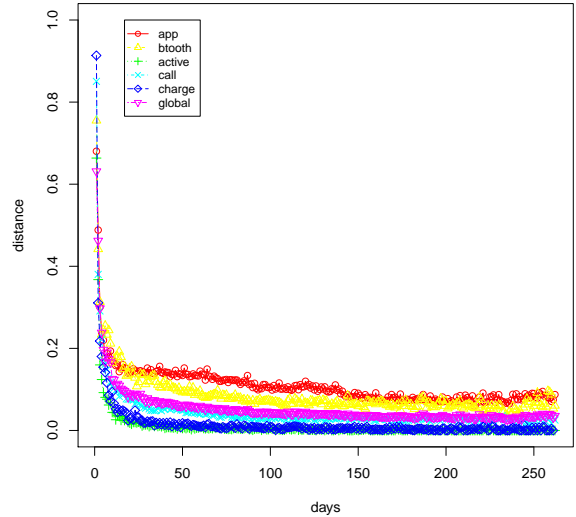
(b) Temporal models

Fig. 5. Comparison of spatial and temporal models between current day and preceding day on MIT dataset. Low distance indicates increased similarity.

the one that emerges after training. Figure 7 shows how the comfort value at $2^{nd}$ percentile changes over time. The values from each user in the data is averaged. Furthermore, Figure 6 shows comfort values at $2^{nd}$ percentile for each user, which we also use for our computations. In all three datasets, the comfort level increases substantially in the first few weeks. Given a 1 minute sampling rate, approximately 1440 comfort scores are computed per day. We compute the comfort level at $2^{nd}$ percentile every day for each user. The implication of $2^{nd}$ percentile is that, as long as the new data follows the distribution captured by the model, approximately 98% of the data will be above the threshold and thus will not result in explicit authentication.

While the value at percentile provides a data-driven way to set the threshold, the user can select a suitable percentile for his/her requirements (and one could imagine that this value is set at default, and perhaps updated by the phone manufacturer or service provider). For example, if the user keeps the percentile at 2, he/she should expect explicit authentication to kick in roughly 2% of the time. If this is not acceptable from a (usability) security standpoint, he/she can (decrease) increase the percentile to accommodate (less) more explicit authentication. This will allow user to interact with the authentication scheme (in the form selecting a point within a range) without having to set a specific threshold. Though in reality, the user may only have insight into the usability impact of setting the threshold. We discuss issues related to security and impersonation below in Section V-C.

## V. EVALUATION

This section presents case studies of legitimate and attack usage scenarios to investigate how well our implicit authentication technique might perform in practice.

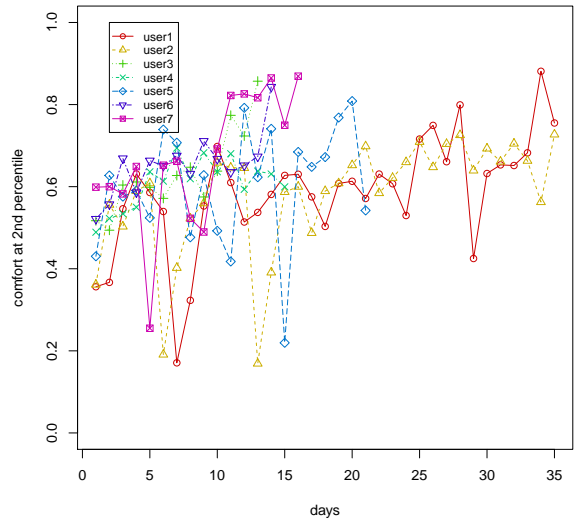

Fig. 6. Comfort score at $2^{nd}$ percentile over time for each user in GCU dataset.

### A. Long term study of comfort

Rice and MIT datasets contain user behaviour over 6 months. Additionally, two users in the GCU dataset collected data for over 6 months and they were asked to keep detailed diaries of their daily lives during this long term experiment. Figure 8 shows the comfort levels devices observe from a user from each dataset during the long term experiment. Other users in the datasets exhibit similar properties.

Figure 8 shows that devices establish different comfort levels for different users. For GCU User 2, the typical comfort was around 0.5, whereas for the selected MIT and Rice users, it was closer to 0.2. This is mainly due to two reasons: (i) each dataset consists of different sets of sensors, which may provide different comfort levels, and (ii) each user may have a different level of predictability in their lives. Therefore, it is crucial to establish comfort level and detection thresholds per user since there is no one size-fits-all solution.

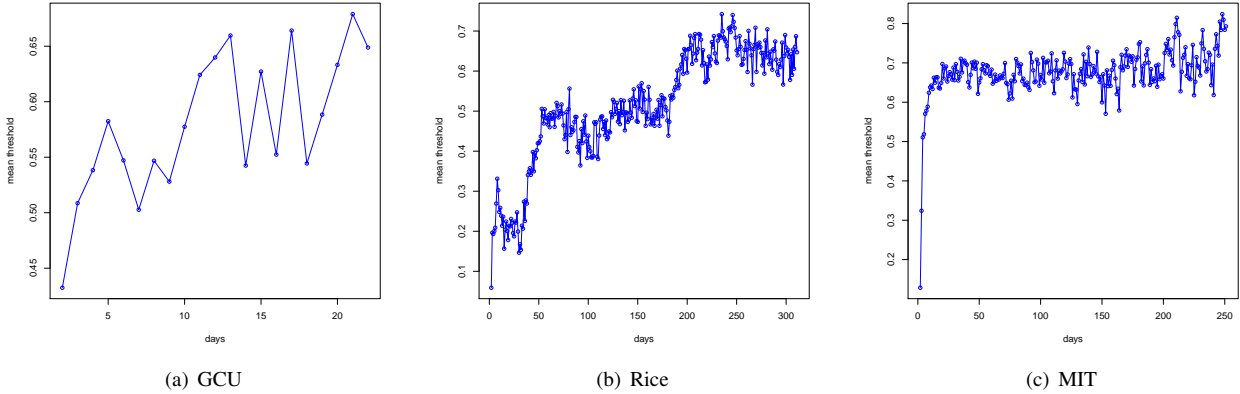

Fig. 7. Comfort score at $2^{nd}$ percentile over time. Score is averaged over all users per dataset.

Another phenomenon observable from Figure 8 is *behavioural drift*. In long-term studies, we observed a slow and gradual decrease in overall comfort as time passes. After three weeks of training, GCU User 2's device encounters comfort levels above 0.7, which slowly drops down to 0.5 after 6 months. Similarly, Rice User A08 starts out with comfort levels of 0.5 which drops below 0 after one year. On the other hand, MIT User 4 encounters comfort levels up to 1.0 which drops to 0.5 in six months. This supports our argument for retraining: as the device recognises behaviour drift, it should go back to training mode again until the new behaviour is learned. Our technique provides a suitable means to detect behaviour drift by checking the distribution of comfort levels daily and suggesting to the user to put the device in training mode, if the distribution varies sufficiently from that previously established, for a period of time. Needless to say, to prevent devices from learning malicious behaviour, the transition from deployment back to training occurs after explicit authentication and only for a brief period of time, such as a few hours, after which the user needs to authenticate again to update the model.

Note that we use the term 'retraining' since our current model is based upon a fixed profile that is used during deployment. Drift from this profile can automatically trigger periodic retraining. An alternative would be to use a more dynamic profile that gradually changes according to a user's behaviour drift. However, we believe that such a model may not be as effective since more frequent updates could require user confirmation, and our goal is to reduce the frequency of explicit user interaction. In addition, our experimental data suggests that behavioural drift might not be so noticeable till after a period of several months (e.g., 6 months for some users). Thus, the relatively stable behaviour of users over time suggests that a fixed model, with periodic retraining, is sufficient for our purposes.

### B. Changes in behaviour

The long term study of comfort levels on the three datasets shown in Figure 8 shows behavioural drift in which the comfort level decreases over time. This can cause an increase in the number of explicit authentication requests if the comfort drops below the detection threshold frequently. Over the period of a few months, the comfort level drops substantially for all three users.

To investigate the effect of moving to a new city, we plot the comfort level observed for GCU User 2 in Figure 9. On June 15, the user moves to a different city which causes the comfort level to drop. In Figure 9(a) the device uses the old model after the move resulting in a low comfort level. Figure 9(b) shows the change in comfort level when the user choses to update the existing model whereas in Figure 9(c), the old model is scrapped and a new model is trained. Results show that re-training the existing model 9(b) is comparably effective as training a new model 9(c) when the established routine changes, although retraining the existing model appears to provide higher comfort scores for the first few days.

### C. Attack Case Studies

Four different attack scenarios are formulated based on the attacker's level of access to a user's frequent locations and his/her knowledge about the user's behaviour [3]. To this end, we define two adversarial levels based on their knowledge of the user and the authentication scheme. An *uninformed adversary* knows very little about the user and their routine whereas an *informed adversary* possesses some knowledge of the users, for example, applications are frequently used (e.g., a close friend or family member). Additionally, we define an *outsider* to be a person who steals the device and runs away. On the other hand, an *insider* has access to a location that user frequently visits and attempts to use the device at a location familiar to the device (e.g., an office mate).

To facilitate our experiments, we asked one of our lab members to use their device until the profile settled. After the profile settled, the device owner was asked to use the device normally to establish the typical comfort level. We then devised four scenarios, detailed as follows:

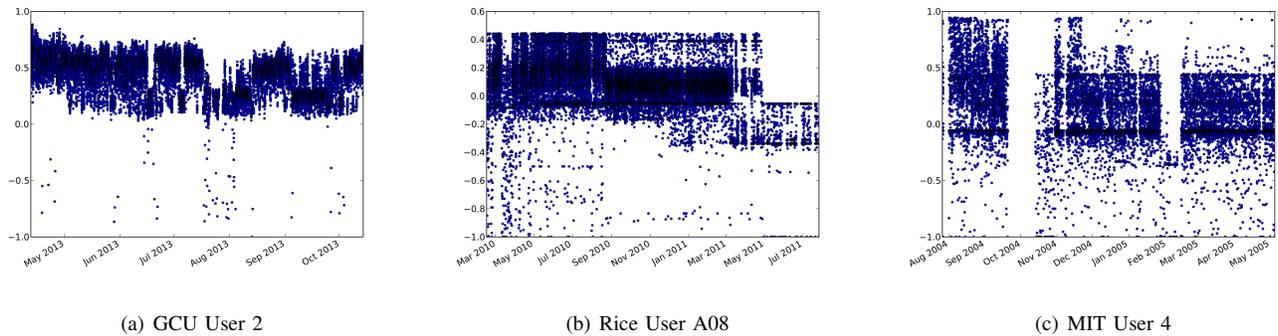

(a) GCU User 2
(b) Rice User A08
(c) MIT User 4

Fig. 8. Long term comfort levels for sample users from each dataset. Other users exhibit similar properties.

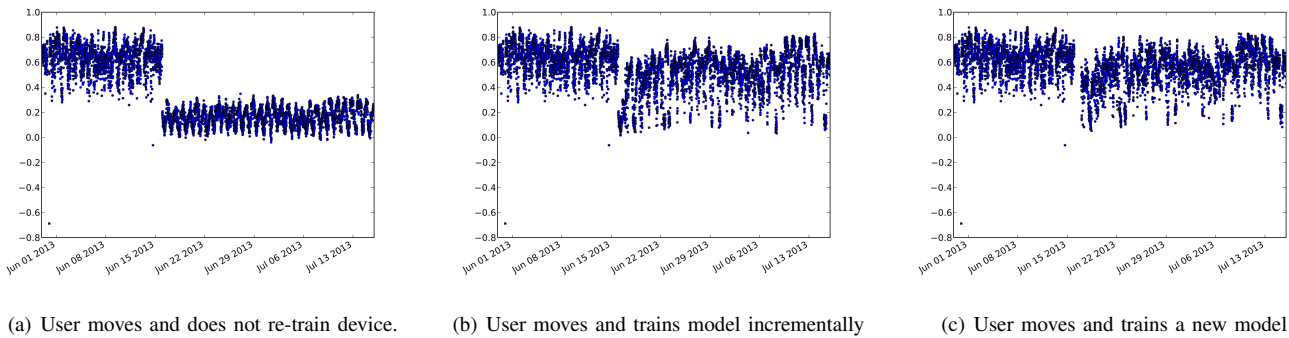

(a) User moves and does not re-train device.
(b) User moves and trains model incrementally
(c) User moves and trains a new model

Fig. 9. The change in comfort level when user moves with and without training.

**Uninformed outsider:** The attacker took device and carried it with them for one day. The attacker lived in another city, ensuring that the locations they frequent were different from the owner's. The attacker was provided no additional information on how the owner uses their device. The attack started at 2pm. This is similar to a typical device theft.

**Informed outsider:** The attacker took the device and carried it with them for one day. The attacker's work and home did not intersect with the device owner's, ensuring that locations are different. However, the attacker was provided with a list of applications that the owner uses. The attack started at 3pm. This attack corresponds to a device theft scenario where the attacker tries (albeit with limited sophistication) to evade the authentication technique.

**Uninformed insider:** The attacker took the device to a cafeteria that the device owner goes to for lunch breaks. Thus, the location was known to the device but was not one of the frequent locations such as home. The participant attacker was provided no additional information on how the owner uses their device. The attacker used the device between 1pm and 5pm. This scenario corresponds to an insider attack in which a naive insider attempts to use the device at a reasonably known location.

**Informed insider:** The attacker, who was the owner's housemate, used device at the owner's home for the day. Thus, the location was well known to the device. The participant attacker was provided with additional information on how the owner uses their device as well. The attacker used the device between 1pm and 5pm. This scenario corresponds to an insider attack in which a capable insider attempts to use the device at a well-known location.

Figure 10 shows the changes in comfort levels for attack scenarios. The horizontal line shows the detection threshold and vertical lines indicate the time of the attack. All four attacks caused comfort level to drop considerably, hence setting off explicit authentication. Informed attacks produce higher comfort levels compared to uniformed attacks but not enough to bypass detection.

After deployment, the device automatically determined the detection threshold to be 0.2. Therefore during the attack scenarios, if the comfort level drops below 0.2, the device requires explicit authentication to unlock. Table III provides a summary of results for the attacks as well as the control sample day during normal use. Results are provided in terms of the average comfort, percentage of events that are flagged as attacks and the time it took for the device to lock down. The results indicate that uninformed attacks, both outsider and insider, are detected quicker than the informed attacks. Additionally during all four attacks, the comfort level dropped substantially below the owner's average. With the exception of informed insider attacks, arguably the most sophisticated of the four, attacks were detected with over 95% detection rate. Even when the attacker attempted evasion, the device locked in under 15 minutes.

Of course, this initial security analysis only demonstrates the relative difference between our four attack scenarios. Further study is required with a larger sample of users, and with different model parameters.

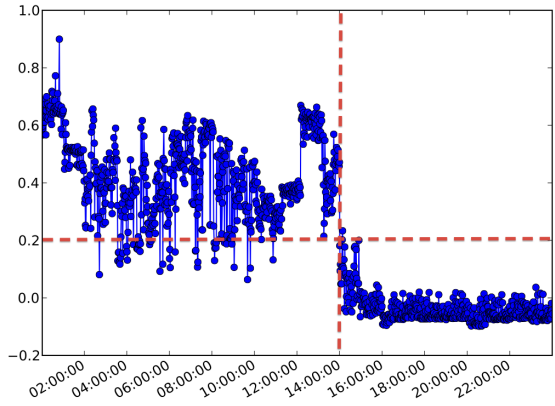
(a) Uninformed Outsider.

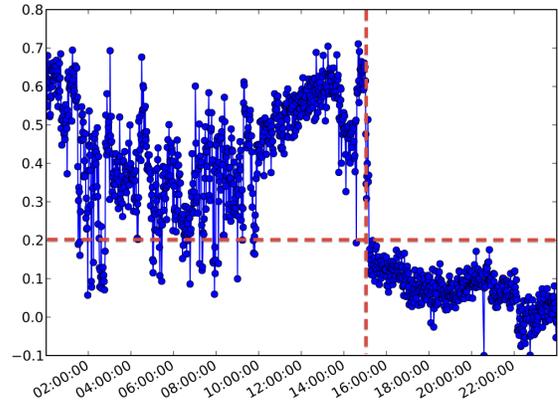
(b) Informed Outsider.

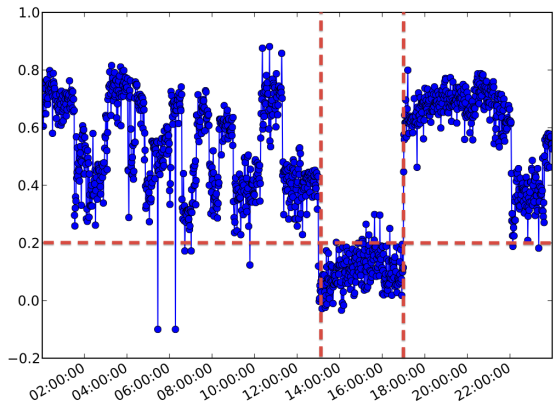
(c) Uninformed Insider.

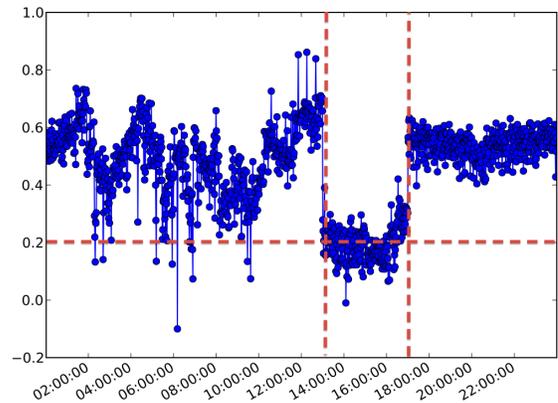
(d) Informed Insider.

Fig. 10. Comfort levels observed during attacks.

TABLE III. AVERAGE COMFORT LEVEL AND THE RATIO OF EVENTS THAT ACTIVATED EXPLICIT AUTHENTICATION.

| Scenario | Comfort (mean) | Detection rate | Time to detect |
|---|---|---|---|
| Owner's sample day | +0.503 | 1.01% | – |
| Uninformed outsider | -0.038 | 99.44% | 122 sec |
| Informed outsider | +0.074 | 98.12% | 851 sec |
| Uninformed insider | +0.102 | 95.85% | 239 sec |
| Informed insider | +0.196 | 53.21% | 717 sec |

## VI. CONCLUDING REMARKS

In this paper, we study the use of sensor-based authentication on mobile devices. To this end, we proposed a lightweight profiling technique that is based on probability density functions of sensor data that provide temporal and spatial awareness. Given that building probability density functions take near linear time, we believe it is a suitable method for performing on-device training. The training process and determination of a suitable threshold is data driven and automated so as to reduce costly human involvement as much as is possible.

While our results are derived from our new model, our broader goal was to begin to quantify the effectiveness of sensor-based authentication. To this end, it is crucial to establish the operational advantages and limits of sensor-based authentication and to build more robust models that account for slight changes in user behaviour while being able to rapidly detect truly malicious behavioural changes, all while not adversely impacting device resources (e.g., battery).

Our future work includes several investigations along these lines. For example, we are expanding our analysis to incorporate various supervised learning techniques for profiling. We are also investigating the effectiveness of different model parameters, and of different sensors (perhaps in different situations or for different users), that might lead to a weighted comfort computation. Finally, while we used a sampling rate of 1 minute for our experiments, we are also investigating the use of adaptive sampling, at least to reduce the battery consumption of our techniques.

ACKNOWLEDGMENT

Thanks to the MoST reviewers for their helpful comments. The research leading to these results has received funding from the People Programme (Marie Curie Actions) of the European Union's Seventh Framework Programme (FP7/2007-2013) under REA grant agreement no PIIF-GA-2011-301536.